\documentclass[12pt,letterpaper]{article} \usepackage[latin1] {inputenc} 
\usepackage{amsmath} \usepackage{amsfonts} \usepackage{amssymb} 

\begin{document}
\title{Lagrangian and Noncommutativity}
\author{Ignacio Cortese and  J. Antonio Garc\'\i a\footnote{email: garcia@nuclecu.unam.mx}\\
\em Instituto de Ciencias Nucleares, \\
\em Univesidad Nacional Aut\'onoma de M\'exico\\
\em Apartado Postal 70-543, M\'exico D.F., M\'exico}
\maketitle
\abstract{We analyze the relation between the concept of auxiliary variables and the Inverse problem of the calculus of variations to construct a Lagrangian from a given set of equations of motion. The problem of the construction of a consistent second order dynamics from a given first order dynamics is investigated. At the level of equations of motion we find that this reduction process is consistent provided that the mapping of the boundary data be taken properly  into account. At the level of the variational principle we analyze the obstructions to construct a second order Lagrangian from a first order one and give an explicit formal non-local Lagrangian that  reproduce the second order projected dynamics. Finally we apply our ideas to the so called    
``Noncommutative classical dynamics''.}

\section{Introduction}

Inspired from recent results in string theory \cite{Douglas}  some authors, 
have  proposed
a ``noncommutative quantum mechanics on the plane'' \cite{Mezic}. Here by quantum
mechanics they mean a classical field theory for the classical
``probability field'' $\psi(x)$ defined in a space where the coordinates do
not commute
\begin{equation}
\label{nc-space-time}
[x^i,x^j]=i\theta^{ij},
\end{equation}
where $\theta^{ij}$ is a constant antisymmetric matrix. The tools used recently for the analysis of noncommutative field
theory can be applied to this case. In particular, the introduction of
noncommutativity in field theories renders the classical field functions
on space-time to operators or matrices whose treatment resemble very
nearly the techniques usually employed to quantize classical field
theories \cite{Douglas-Nekrasov}. It is also possible  to use the Weyl or Moyal star product in a 
constructive way to ``deform'' the standard product of functions. 
Indeed, a procedure to obtain, in the appropriate limit, this non-commutative quantum mechanics starting from a noncommutative field theory was given by \cite{Ho-Kao}.
The complete picture that emerges from here is quite non-local and has  been investigated only for some simple cases \cite{Gamboa}. 

Another beautiful approach to obtain a consistent quantum dynamics in phase space with a curved symplectic structure, in contrast to the noncommutative space-time (\ref{nc-space-time}), is deformation quantization \cite{defor}. Here we start
with an associative commutative algebra $\cal A$ and deform it
 into a new associative but noncommutative new algebra. This can be
done using $\hbar$ as a deformation parameter but also we can use $\sigma^{ij}$ as a
deformation parameter (see for example \cite{Dayi}) where
\begin{equation}
\label{PB}
\{z^i,z^j\}=\sigma^{ij}.
\end{equation}
Here $\{z^i,z^j\}$ denotes the Poisson bracket in the ``phase space'' defined by the coordinates $z^i$ and  $\sigma^{ij}$ is a constant matrix. In this case, we can use an ordering prescription, say symmetric Weyl ordering, to get a consistent picture of the algebra of observables by replacing the standard product of operators by the star or Moyal product in the space of its associated symbols \cite{Harvey}. The picture that emerges from here is a quantum mechanics on phase space defined by some Hamiltonian and the structure given by (\ref{PB}). 
Of course this approach is very different from the one mentioned above.
To recover from deformation quantization a Schroedinger equation and a wave function in configuration space, when the coordinates do not commute, is not an easy task. The reason is that the usual momenta associated with the noncommutative coordinates are not  auxiliary variables, in the usual sense. The consequences that this fact has upon the resulting quantum mechanics are quite severe. Indeed, it is not known how to project, in a consistent way,  the Wigner distribution functions associated with a given problem to distributions of probability in configuration space. 
For non constant $\sigma^{ij}$ the analysis is much more involved. Very interensting recent results along this line started with the seminal work \cite{Konsev} and have been applied to a sigma model in \cite{Cattaneo-Felder}.

One obvious question is if there exists a noncommutative classical
mechanics that upon some quantization procedure, say canonical
quantization, gives the ``noncommutative quantum mechanics on the plane''. 
Since Classical Mechanics (in its first order formulation) is covariant with respect to a wide class of transformations --Darboux transformations-- the question of noncommutativity in Classical Mechanics does not make sense unless we fix coordinates using some physical criteria.

An interesting approach not pursued here could be to consider classical dynamics with a noncommutative time. The result of such analysis could be similar to the non-local classical dynamics analyzed in \cite{Gomis-Llosa}. 

Here we will consider a different problem which is not related, at least in any direct way, to the approach of constructing noncommutative field theories\footnote{ A recent attempt to construct a noncommutative field theory with fields that are noncommuting can be found in \cite{Gamboa-Mendez}}. The approach that we will present here  could be related with deformation quantization.
The first problem, at a Classical level, is how to define in a consistent way the configuration  space dynamics when the coordinates on this space do not  commute in the symplectic sense (\ref{PB}). Namely, to define from a given first order variational principle the associated second order variational principle in configuration space upon the elimination of momenta from the first order dynamics.
As far as we know, there is not a complete answer to this problem. A previous analysis of the Inverse problem from this perspective can be found in \cite{Henneaux}.

In its simplest setting the problem can be  stated as follows: giving a classical system described by some Hamiltonian
of the form $H=T+V$, and some (constant) symplectic structure, can we define the
second order equations of motion {\em in configuration space} in such a way that to every solution of the original first order system of differential equations we can associate a solution of the second order equations in configuration space?
And, as a second step, can we define the associated quantum theory?

A previous attempt to analyze the problem of the quantization of classical systems with nonstandard symplectic structures can be found in \cite{Hojman-Shepley}, where the importance of the existence of a second order Lagrangian for the quantization is remarked. In this context  a discusion about the restricted inverse problem of the calculus of variations and the quatntization of first order systems can be found in 
\cite{Acatrinei}.

The aim of this note is to give an answer to the first of these 
questions and to show
that there are still some uncovered problems.
The analysis of the second question  will be presented elsewhere.

For that end we will review
in section 2  the basic ideas of auxiliary variables in first order systems and
the inverse problem of the calculus of variations, i.e, the necessary
and sufficient conditions for the existence of a Lagrangian for a
given set of first order dynamical equations of motion \cite{Hojman-Urrutia}. This problem
is of relevance because there is a relation between the existence of auxiliary variables in the first order dynamics and the existence of a Lagrangian function 
for the configuration space equations of motion. Here we want to confront the following paradox: From one hand a Lagrangian for an integrable system of first order differential equations always exist \cite{Hojman-Urrutia}. From the other hand we can always associate to a given system of second order differential equations a system of first order differential equations. By eliminating the auxiliary variables from the first order system we can recover the second order system and by the same procedure obtain also a Lagrangian for these second order equations, using the first order variational principle. But it is well known that the Lagrangian for the second order differential equations does not always exist \cite{Anderson}.
So, which are the general conditions over the first order Lagrangian function to give, upon reduction, a second order Lagrangian associated to an equivalent second order system?
As we will see, this problem  can be formally solved  with a non-local Lagrangian that gives an equivalent set of equations of motion in the reduced space. The Lagrangian equations of motion in the reduced space are equivalent to a set of second order equations of motion, but the equivalence with the original first order dynamics can be implemented only through a proper consideration of the boundary data and its ``projection'' to the second order dynamics.  
  
Given the coordinates of the configuration space we will investigate in section  3
the case when we have at hand a Hamiltonian to describe the first order dynamics. In section 4 we will present a procedure to construct a non-local Lagrangian for theories with a given Hamiltonian of the form $H=T+V$ and a constant symplectic structure.
In section 5 we will define the ``noncommutative classical dynamics'' . We will analyze the precise equivalence between ``Newtonian equations of motion'' and the first order formulation of the theory. Section 6 is devoted to examples and section 7 to conlusions and perspectives.

\section{First order Lagrangians and the inverse problem of the
  calculus of variations}

In this section we will review two central ideas relevant for our discussion. The first one is the definition of auxiliary variables for a given variational principle \cite{Henneaux-Teitelboim}. The second one is about the existence of a Lagrangian for a given first order system of differential equations \cite{Hojman-Urrutia}. Then we will relate this two ideas to construct a Lagrangian for the reduced system that results from the elimination of the auxiliary variables in the original system and point out the possible obstructions to perform such procedure. 

\subsection{Auxiliary variables}

Consider the action
\begin{equation}
S=\int dt L(x,\dot x, \ddot x,....),
\end{equation}
where the Lagrangian depends on the coordinates $x^a, a=1,2...N$ and its derivatives with respect to time up to any finite order. The equations of motions will be denoted by
\begin{equation}
\frac{\delta L}{\delta x}=\sum_{n=0}(-1)^nD^n\frac{\partial L}{\partial  x^{(n)}},
\end{equation}
where $D$ is the total derivative with respect to time given by
\begin{equation}
D=\dot x\frac{\partial }{\partial x}+\ddot x\frac{\partial }{\partial \dot x}+...+\frac{\partial }{\partial t},
\end{equation}
and $x^{(n)}$ denotes the $n-$derivative with respect to time of the coordinate $x$. Now suppose that the coordinates can be splited into two groups
\begin{equation}
x^a\to (y^i,z^\alpha),
\end{equation}
such that
\begin{equation}
\label{desp-z}
\frac{\delta L}{\delta z^\alpha}=0 \Leftrightarrow z^\alpha=Z^\alpha(y^i,{\dot y}^i, {\ddot y}^i, ...).
\end{equation}
Here we want to stress that the procedure to obtain the variables $z^\alpha$ in terms of the reduced space variables $y^i$ is purely algebraic\footnote{ There are some particular examples where this step can be realized by integration. As far as we know there is no a systematic generalization of the ideas below for this case.}. In that case the variables $z^\alpha$ are called {\em auxiliary variables}. The reduced Lagrangian $L_R$ is defined by
\begin{equation}
L_R(y^i,{\dot y}^i, {\ddot y}^i,....)=L(y^i,Z^\alpha,{\dot y}^i,{\dot Z}^\alpha, ....).
\end{equation}
Then we can state the following Theorem: The original equations of motion
\begin{equation}
\frac{\delta L}{\delta y^i}=0,\qquad \frac{\delta L}{\delta z^\alpha}=0,
\end{equation}
are equivalent to the equations
\begin{equation}
\frac{\delta L_R}{\delta y^i}=0,\qquad
z^\alpha=Z^\alpha.
\end{equation}
The proof is straightforward using the chain rule
\begin{equation}
\label{chain-rule}
\frac{\delta L_R}{\delta y^i}=\frac{\delta L}{\delta y^i}+\frac{\delta L}{\delta z^\alpha}\frac{\partial Z^\alpha}
{\partial y^i}-\frac{d}{dt}\left(\frac{\delta L}{\delta z^\alpha}\frac{\partial Z^\alpha}
{\partial {\dot y}^i}\right)+....(-1)^k\frac{d^{(k)}}{dt^{(k)}}\left(\frac{\delta L}{\delta z^\alpha}\frac{\partial Z^\alpha}
{\partial  {y^{(k)}}^i}\right).
\end{equation}
The equations  $\frac{\delta L}{\delta z^\alpha}=0$ imply $z^\alpha=Z^\alpha$ (by asumption) and $\frac{\delta L_R}{\delta y^i}=0$ (by (\ref{chain-rule})). Conversely,
the equations $\frac{\delta L_R}{\delta y^i}=0$ and $z^\alpha=Z^\alpha$ imply $\frac{\delta L}{\delta z^\alpha}=0$ (by assumption) and $\frac{\delta L}{\delta y^i}=0$ (by (\ref{chain-rule})).

As an example lets see how this works with the standard Hamiltonian dynamics. The first order Lagrangian is
$$
L=\dot x p -\frac12 p^2 -V(x),
$$
while the reduced Lagrangian (after the elimination of the momenta)
is
$$
L_R=\frac12 {\dot x}^2-V(x).
$$
To see the equivalence we write the chain rule (\ref{chain-rule})
$$
\ddot x+V'(x)=\frac{\delta L}{\delta x}-\frac{d}{dt}\frac{\delta L}{\delta p},
$$
which shows explicitly the required equivalence. It is interesting to notice that if we fix as initial conditions $x,p$ at some initial time $t_0$ in the equations of motion associated to $L(x,p)$ then the initial conditions for the reduced system $\ddot x+V'(x)=0$ are the standard Newtonian position and velocity at initial time $t_0$.

\subsection{Inverse problem for the calculus of variations for first order systems and reduction}

In this subsection we will construct a criteria for the existence of a second order Lagrangian given a first order one based on the concept of auxiliary variables.

By splitting the original variables $x^a$ into $z^\alpha$ and $y^i$ and by choosing the $y^i$ as the variables that define the reduced space we can write a relation of the form
\begin{equation}
\label{GIS}
G_i\left(\frac{\delta L}{\delta x^a},  
\frac{d}{dt}(\frac{\delta L}{\delta x^a})\right)=M_i(\ddot y^j,{\dot y}^j, y^j),
\end{equation}
where $G_i$ are linear combinations of the first order equations of motion associated to the original first order Lagrangian $L$ and its derivatives. The question is if this linear combinations can be written as a variational derivatives with respect to the reduced variables $y^i$ for some function $L_R(\dot y, y)$, {\em i.e.}, if the equations of motion of the reduced space $M_i(\ddot y, \dot y ,y)=0$ comes from a variational principle. This will be true iff
\begin{equation}
\label{Helmholtz}
\frac{\delta M_i(t)}{\delta y^j(t')}=\frac{\delta M_j(t')}{\delta y^i(t)}.
\end{equation}
These conditions are the Helmholtz conditions for the restricted inverse problem of the calculus of variations associated to the system $M_i=0$. If this conditions are meet then $G_i$ are  variational total derivatives. A generalization of this problem is the question about the existence of a matrix $\Lambda^{ij}(y,\dot y) $ with $\det\Lambda\not=0$ such that the equivalent  system $\Lambda^{ij}M_j=0$ can be deduced from a variational principle, i.e., if an ``integrating factor'' exists such that the relations (\ref{GIS}) are total variational derivatives. This is the unrestricted inverse problem of the calculus of variation. A  review about this interesting problem can be found in \cite{Anderson}.

The conditions (\ref{Helmholtz}) can be written in terms of the original first order Lagrangian $L$ through the functions $G_i$ to see which are the conditions under the Lagrangian $L$ to give a second order variational formulation of the system $M_i$ upon the reduction process. 

If the variables used in the reduction process are auxiliary then $G_i$ is a total functional derivate and the conditions (\ref{Helmholtz}) for the existence of a Lagrangian are automatically satisfied. The reduced Lagrangian $L_R$ is a Lagrangian for the reduced second order equations of motion. 

The functions $G_i$ contains the precise information about the algebraic manipulations performed on the original system of equations to arrive at the second order ones in the reduced space. The relations (\ref{GIS}) do not imply that the two sets of equations 
$\frac{\delta L}{\delta y^i}=0, \frac{\delta L}{\delta z^\alpha}=0$ and 
$M_i=0, \frac{\delta L}{\delta z^\alpha}=0$ are equivalent. The equivalence between these two sets of equations depends crucially on the structure of the original equations of motion and on the reduction process. There are two interesting scenarios: a) reduction with auxiliary variables while maintaining the equivalence (as in the standard Hamilton to Lagrange description in Classical Mechanics), b) reduction with non-auxiliary variables while maintaining the equivalence (a case where the second order Lagrangian does not exist (in the restricted sense stated above)). 
 
If the variables used in the reduction are not auxiliary $G_i$ is not a total variational derivative and the Lagrangian associated to the system $M_i$ does not exist. But it is still possible that a Lagrangian associated with the equivalent system $\Lambda^{ij}M_j=0$ 
could be constructed. We will not pursue this general problem here but we will take a shortcut to construct a non-local Lagrangian in the case where we do not have auxiliary variables to perform the reduction procedure and the symplectic structure is constant.

The analysis of the unrestricted inverse problem of the calculus of variations for first order system is, in contrast, much more easy to workout. It can be stated as follows \cite{Hojman-Urrutia}: Given a first order system of differential equations in a space defined by the coordinates $x^a, a=1,2...2n$
\begin{equation}
\label{fo-eq}
{\dot x}^a-f^a(x)=0,
\end{equation}
we can construct a Lagrangian for this given set of equations in the form
\begin{equation}
\label{eq-mot-fo}
\frac{\delta L}{\delta x^a}=\sigma_{ab}(x)({\dot x}^b-f^b),
\end{equation}
where $\sigma_{ab}$ is a non singular antisymmetric matrix and $\det\sigma\not=0$. This Lagrangian can be constructed by the following procedure \cite{Hojman-Geo}. First write a general Lagrangian function as
\begin{equation}
\label{fo-lag}
L=\ell_a(x) ({\dot x}^a-f^a).
\end{equation}
Now, using (\ref{eq-mot-fo}) we see that $\ell_a(x)$ must satisfy
\begin{equation}
\label{lie}
{\cal L}_f \ell_a=\frac{\partial\ell_a}{\partial x^b}f^b+ \frac{\partial f^b}{\partial x^a}\ell_b=0,
\end{equation}
where ${\cal L}_f$ is the Lie derivative along the vector field $f^a$ and the matrix $\sigma$ is given by
\begin{equation}
\label{sigma}
\sigma_{ab}=\frac{\partial\ell_b}{\partial x^a}-\frac{\partial\ell_a}{\partial x^b},
\end{equation}
in order that the equations of motion associated with the Lagrangian (\ref{fo-lag}) be equivalent to the set of the given equations (\ref{fo-eq}). The condition (\ref{lie}) always has solution (at least locally) \cite{Hojman-Urrutia} so it is always possible to construct a variational principle for a given set of first order differential equations.

In particular, a Hamiltonian form for the given system (\ref{fo-eq}) can be constructed if we can find a function $H(x)$ such that
\begin{equation}
\label{Ham-cond}
\ell_a f^a=H(x),\qquad \sigma_{ab} f^b=\frac{\partial H}{\partial x^a}.
\end{equation}
Conversely, given $H(x)$ satisfaying (\ref{Ham-cond}) the function $\ell_a$ satisfy (\ref{lie}). The symplectic structure of the  resulting Hamiltonian system is given by $\sigma^{ab}$ ($\sigma=d\ell$). 

We can always choose coordinates (at least locally) such that the given system (\ref{fo-eq}) can be transformed to the ``normal form''
\begin{equation}
\label{change-variable}
{\dot x}^a-f^a(x)=0 \rightarrow {\dot \xi}^i-\xi^{i+n}=0,\quad {\dot \xi}^{i+n}-f^{i+n}(\xi)=0.
\end{equation}
Using these variables and taking as the reduced space $\xi^i$ we obtain, upon reduction, the second order system
\begin{equation}
\ddot \xi^i-F^i(\xi^j,\dot \xi^j)=0, \qquad F^i(\xi^j,\dot \xi^j)\equiv f^{i+n}(\xi)\Big|_{\xi^{i+n}={\dot \xi}^i}.
\end{equation}

The Lagrangian equations of motion (\ref{eq-mot-fo}) take the following form
\begin{eqnarray}
\label{eq-mot-xi}
\frac{\delta L}{\delta \xi^i}=\sigma_{ij}({\dot \xi}^j-\xi^{j+n})+\sigma_{ij+n}({\dot \xi}^{j+n}-f^{j+n}(\xi))=0,\\
\label{eq-mot-xi+n}
\frac{\delta L}{\delta \xi^{i+n}}=\sigma_{i+nj}({\dot \xi}^j-\xi^{j+n})+\sigma_{i+nj+n}({\dot \xi}^{j+n}-f^{j+n}(\xi))=0.
\end{eqnarray}
The variables $\xi^{i+n}$ are auxiliary variables iff $\sigma_{i+nj+n}=0$. We can define the second order Lagrangian for the reduced system 
by applying the reduction procedure outlined in section 2.
In the case where $\sigma_{i+nj+n}\not=0$ the variables $\xi^{i+n}$ are not auxiliary and the reduction process can not help to construct a Lagrangian for the second order system. We will see below how to use the change of variables (\ref{change-variable}) to construct an equivalent Lagrangian associated to the reduced original system of second order equations of motion. The two conditions upon the transformation  (\ref{change-variable}) are 1) $\sigma_{i+n j+n}(\xi)=0$ and 2) that the transformation be symmetry of the equations of motion.

Defining the configuration space by the variables $\xi^i$ we can construct the explicit relation between the equations of motion associated to the first order system and the reduced space equations  (\ref{GIS}). It is easy to write this relation using the matrix $\sigma$. The result is
\begin{equation}
\label{GIS-sigma}
\sigma^{i+n j}\frac{\delta L}{\delta \xi^{j}}+
\sigma^{i+n j+n}\frac{\delta L}{\delta \xi^{j+n}}
+\frac{d}{dt} \Big(\sigma^{ij} \frac{\delta L}{\delta \xi^{j}} +\sigma^{ij+n}  \frac{\delta L}{\delta \xi^{j+n}}\Big)
={\ddot \xi}^k-f^{k+n}(\xi),
\end{equation}
where $\sigma^{ij}$ is the inverse matrix of the matrix $\sigma_{ij}$. This relation suggest that the combinations of the Lagrangian equations of motion
\begin{equation}
\sigma^{i+n j}\frac{\delta L}{\delta \xi^{j}}+
\sigma^{i+n j+n}\frac{\delta L}{\delta \xi^{j+n}}=\dot \xi^{i+n}-f^{i+n}(\xi),
\end{equation}
and 
\begin{equation}
\sigma^{ij} \frac{\delta L}{\delta \xi^{j}} +\sigma^{ij+n}  \frac{\delta L}{\delta \xi^{j+n}}={\dot \xi}^i-\xi^{i+n},
\end{equation}
are, in fact, equivalent to the system
\begin{equation}
{\dot \xi}^i-\xi^{i+n}=0, \qquad \dot \xi^{i+n}-f^{i+n}(\xi)=0,
\end{equation}
iff $\det\sigma\not=0$.

But this equivalence is not ``variationally admissible'' \cite{Novikov} because does not define auxiliary variables in the reduction process. This imply that the Lagrangian for the second order system (``Newton equations'') can not be defined by using the first order Lagrangian.

Notice that here $\xi^i$ and $\xi^{i+n}$ do not have the meaning of coordinates and momenta, respectively. The nessesary conditions for the existence of a first order Lagrangian with auxiliary variables $\xi^{i+n}$ are  the condition (\ref{lie}) in terms of the new variables $\xi$ and $\sigma_{i+n j+n}(\xi)=0$.

For example consider the isotropic harmonic oscillator in the plane with mass and frequency equal to one. The phase space is defined by $q_1,q_2,p_1,p_2$ with the standard symplectic structure. Now consider the first order Lagrangian \cite{Hojman}
\begin{equation}
\label{ejos}
L=E(q_i{\dot p}_i-p_i{\dot q}_i)-H
\end{equation}
where $E=\frac12\sum_i(q_i^2+p_i^2)$ and $H=-2E^2$. The matrix $\sigma_{ab}$ is
\begin{equation}
\sigma_{ab}=\begin{pmatrix}
p_iq_j-q_ip_j&2 E\delta_{ij}+q_iq_j+p_ip_j\\
 -(2 E\delta_{ij}+q_iq_j+p_ip_j)& p_iq_j-q_ip_j
\end{pmatrix}.
\end{equation}
The momenta $p_i$ are not more auxiliary variables and the reduction process fails. So we can not define the reduced second order Lagrangian associated to the first order Lagrangian (\ref{ejos}) by the process of elimination of the variable  $p_i$ by using its own equations of motion. Nevertheless, the dynamical system described by (\ref{ejos}) is equivalent to the original harmonic oscillator. Notice that the phase space in this description is highly noncommutative and that a second order Lagrangian for the equivalent dynamics, i.e., the harmonic oscillator, of course, can be constructed. Another examples along these lines can be found in \cite{Stichel}.

We can use Darboux transformations to construct a Lagrangian for a given set of second order equations. The procedure is: 1) Choose the variables that we want to eliminate, 2) Construct a Darboux transformation that renders those variables to auxiliary variables, 3) Eliminate those variables from the first order system and the Lagrangian.
At the end we will have a Lagrangian for a system of second order equations of motion. If this equations are equivalent to the original ones we have also an equivalent Lagrangian for the given equations. The requirement upon the Darboux transformation is that its projection to configuration space (that results in a function of coordinates and its derivatives) be a symmetry of the equations of motion, i.e.,  ${\cal M}^i=\Lambda^i_j M^j$ where $M^i$ denotes the original set of equations of motion and ${\cal M}^j$ the transformed one and $\Lambda^i_j$ a regular matrix. 

\section{Hamiltonian Systems}

In this section we will analyze the case when we have at hand a Hamiltonian function to describe the given system of first order differential equations. In this case as we noticed, the matrix (\ref{sigma}) corresponds to the symplectic structure of the Hamiltonian system. So we will analyze the role played by the matrix $\sigma_{ij}$ in the definition of the Poisson bracket in the associated ``phase space''. In particular, we will see that the matrix $\sigma_{ij}$ is related with the noncommutativity for the reduced coordinates (``configuration space''). In the case
where $\sigma_{ij}\not =0$ we will talk about a ``noncommutative configuration space''. For that end we will split the variables $x^a$ into $x^i,p_j$. The matrix $\sigma_{ab}$ can be written in the form 
\begin{equation}
\label{sigma-abc}
\sigma_{ab}=\begin{pmatrix}
B_{ij}&-A_{ij}\\
 A^T_{ij}&C_{ij}
\end{pmatrix},
\end{equation}
where $B_{ij}, C_{ij}$ are antisymmetic matrices that depend on the form of the specific Lagrangian that we  choose to describe the given first order system. 
The matrix $\sigma_{ab}$ is by construction antisymmetric, regular $\det \sigma\not=0$ and closed ($\sigma=d\ell, d\sigma=0$ where $d$ is the exterior derivative). So it can be used to construct a Poisson bracket
\begin{equation}
\{F(x),G(x)\}=\frac{\partial F}{\partial x^a}\sigma^{ab} \frac{\partial G}{\partial x^a}
\end{equation}
where $\sigma^{ab}$ is the inverse of the matrix $\sigma_{ab}$. By construction this Possion bracket satisfies the Jacobi Identity (because $\sigma=d\ell$). The inverse matrix $\sigma^{ab}$ has the form
\begin{equation}
\sigma^{-1}=\begin{pmatrix}
{(A^T)}^{-1}CM^{-1}A^{-1}&N^{-1}{(A^T)}^{-1}\\
 -M^{-1}A^{-1}&A^{-1}BN^{-1}(A^T)^{-1}
\end{pmatrix}
\end{equation}
where $M=1+A^{-1}B{(A^T)}^{-1}C$ and  $N=1+{(A^T)}^{-1}CA^{-1}B$. From here we can see the role played by the matrix $C_{ij}$ in the noncommutativity of the reduced coordinates $x^i$. Indeed
\begin{equation}
\{x^i,x^j\}=({(A^T)}^{-1}CM^{-1}A^{-1})^{ij}.
\end{equation}
In this case we say that we have a {\em noncommutative configuration space}.
In particular, when the matrix $A=1$ and $B=0$, this relation becomes
\begin{equation}
\{x^i,x^j\}=C_{ij}.
\end{equation}

For a Hamiltonian system the first order Lagrangian takes the general form
\begin{equation}
L=\ell_a(x){\dot x}^a-H(x),
\end{equation}
and the equations of motion are
\begin{equation}
\frac{\delta L}{\delta x^a}=\sigma_{ab}{\dot x}^b-\frac{\partial H}{\partial x^a}.
\end{equation}
In particular when the Hamiltonian is of the form $H=\frac12 p^2+V(x)$, the equations of motion can be written as
\begin{eqnarray}
\label{eq-mot-x}
\frac{\delta L}{\delta x^i}=B_{ij}{\dot x}^j -A_{ij}{\dot p}_j  -\frac{\partial V}{\partial x^i},\\
\label{eq-def-p}
\frac{\delta L}{\delta p_i}={(A^T)}_{ij}{\dot x}^j + C_{ij}{\dot p}_j  - p_i.
\end{eqnarray}
The case $C=0$ is
\begin{equation}
\sigma^{-1}=\begin{pmatrix}
0&{(A^T)}^{-1}\cr
 -A^{-1}&A^{-1}B{(A^T)}^{-1}
\end{pmatrix},
\end{equation}
and correspond, in view of the analysis of the previous section, to the case where the second order Lagrangian can be constructed because the momenta $p_i$ are auxiliary variables in the usual sense (usual configuration space).
In this case we recover the standard reduction procedure, even if $A,B$ being arbitrary functions of $x^i$. The case $C\not=0$ is more interesting. In this case, the relation analogous to (\ref{GIS}) reads
\begin{eqnarray}
\nonumber
A^{ij}\frac{\delta L}{\delta x^j} + \frac{d}{dt}\left[\frac{\delta L}{\delta p_i}+C_{ik}A^{kl}\frac{\delta L}{\delta x^l}\right]=\\ \label{red-abc}
A^{ij}\left(B_{jk}{\dot x}^k-\frac{\partial V}{\partial x^j}\right)-\frac{d}{dt}\left[C_{ik}A^{kl}\frac{\partial V}{\partial x^l}\right]+\frac{d}{dt}\left[(  (A^T)_{il}+C_{ik}A^{kj}B_{jl}){\dot x}^l\right].
\end{eqnarray}
These equations suggest that the combination
\begin{equation}
\label{p-def}
\frac{\delta L}{\delta p_i}+C_{ik}A^{kl}\frac{\delta L}{\delta x^l}=0,
\end{equation}
play the role of the ``definition of momenta'' for this set of equations. In fact, the original system of equations is equivalent to the system
\begin{eqnarray}
\label{def-p}
C_{ik}A^{kl}B_{lj}{\dot x}^j  -C_{ik}A^{kl}\frac{\partial V}{\partial x^l}+
{(A^T)}_{ij}{\dot x}^j   - p_i=0,\\
\label{dot-p}
B_{ij}{\dot x}^j -A_{ij}{\dot p}_j  -\frac{\partial V}{\partial x^i}=0,
\end{eqnarray}
 iff $\det A\not=0$. By using (\ref{def-p}) as the definition of momenta we can recover the second order dynamics by  pluggin this definition in the equation (\ref{dot-p}).

The reduction is, as in the previous case, not ``variationaly admisible'' so the reduced Lagrangian  can not be constructed by using auxiliary variables.

The second order dynamics is described by the equations of motion
\begin{equation}
\label{sec-or-dy}
A^{ij}\left(B_{jk}{\dot x}^k-\frac{\partial V}{\partial x^j}\right)-\frac{d}{dt}\left[C_{ik}A^{kl}\frac{\partial V}{\partial x^l}\right]+\frac{d}{dt}\left[(  (A^T)_{il}+C_{ik}A^{kj}B_{jl}){\dot x}^l\right]=0.
\end{equation}
These ``Newtonian equations of motion'' must be taken with care because the reduction procedure used to obtain them was not standard. Indeed, a solution to the original first order system with prescribed initial conditions, say by fixing the coordinates and momenta at some initial time by $x^i(t_0)=x^i_0, p_j(t_0)={p_j}_0$, correspond to a solution of the ``Newtonian equations of motion'' (\ref{sec-or-dy}) with the initial conditions
\begin{eqnarray}
x^i(t_0)=x_0^i\\
\Big[ C_{ik}A^{kl}B_{lj}{\dot x}^j  -C_{ik}A^{kl}\frac{\partial V}{\partial x^l}+
{(A^T)}_{ij}{\dot x}^j\Big] (t_0)  = {p_i}_0.
\end{eqnarray}
An interesting observation is that these initial conditions are not the usual initial position and velocities of Newtonian Mechanics but depend on the symplectic matrix $\sigma$ and the dynamics through the potential function $V(x)$. Of course, we can recover the standard case when the matrices $A=1, C=0$.

\section{ A Non local Lagrangian for the reduced system}

In this section we will construct a non local Lagrangian associated with the reduced system (\ref{sec-or-dy}). Notice that we can not construct a standard Lagrangian for this system by the auxiliary variables procedure of our previous section.
We will restrict ourselves to the case when $A,B,C$ are constant matrices. A modification of our procedure in the general case is still an open problem.

We will show here that  in this case, we can construct the equations of motion in the reduced space and its associated Lagrangian. The Lagrangian will be a non local function of the reduced variables and its derivatives with respect to time of any order.   

Observing that we can solve in a formal way the equations of motion (\ref{eq-def-p}) for the ``auxiliary'' variables $p_j$  we have
\begin{equation}
\label{p-j}
 p_j=(1-C\frac{d}{dt})^{jk}(A^T)_{kl}{\dot x}^l,
\end{equation}
where $(1-C \frac{d}{dt})^{jk}$ denotes the inverse of the matrix of $(1-C \frac{d}{dt})_{jk}$.
Notice that we have obtained the variables $p_j$ that are the analogous of the momenta in the standard Hamiltonian formulation in terms of the reduced coordinates $x^j$  and its derivatives with respect to time of infinite order. This relations can be interpreted in the space of solutions as the Green functions associated to the differential equations (\ref{eq-def-p}) given the solution for $x^j(t)$ with some specified boundary conditions. It is possible to extend in some cases the concept of auxiliary variables when we have such differential equations and not the algebraic usual ones as (\ref{desp-z}). As far as we know there is no systematic approach for this case. We will follow here another route. We will take the formal solution (\ref{p-j}) as given and claim that they can be considered as ``auxiliary variables''.

With this observations in mind  we can use (in a formal way) the reduction procedure of section 2. In particular the equations of motion in the reduced space are
\begin{eqnarray}
\label{eq-mot-inv}
A_{ik}\frac{d}{dt}\left[(1-C\frac{d}{dt})^{kl} (A^T)_{lj}{\dot x}^j\right]-B_{ij}{\dot x}^j+\frac{\partial V}{\partial x^i}=0,
\end{eqnarray}
and by multiplication of the matrix $(1-C \frac{d}{dt})_{jk}$ this system is equivalent to the system (\ref{sec-or-dy}). 

To construct the reduced Lagrangian function we observe that the
first order Lagrangian associated to the system (\ref{eq-mot-x},\ref{eq-def-p}) is
\begin{equation}
L=\frac12 \sigma_{ab}x^a{\dot x}^b-H(x),
\end{equation}
where $x^a=(x^i,p_j)$ and the matrix $\sigma_{ab}$ is given by (\ref{sigma-abc}). In terms of $x^i,p_j$ this Lagrangian is
\begin{equation}
L=\frac12 B_{ij}x^i{\dot x}^j+A_{ij}p_j{\dot x}^i+\frac12 C_{ij}p_i{\dot p}^j-H.
\end{equation}
Pluggin $p_j$ from (\ref{p-j}) into this Lagrangian function we obtain the reduced non local Lagrangian
\begin{equation}
\label{nonlocal-lag}
L_{NL}=\frac12 B_{ij}x^i{\dot x}^j +\frac12 A_{ij}{\dot x}^i(1-C \frac{d}{dt})^{jk} A^T_{kl}{\dot x}^l-V(x).
\end{equation}
This reduced Lagrangian function is also of infinite order in time derivatives of the reduced coordinates $x^i$ and to obtain from it the equations of motion (\ref{eq-mot-inv}) we need to apply the extended Euler-Lagrange operator
\begin{equation}
\label{EL-extend}
\frac{\delta L}{\delta x^i}=\sum_{n=0}^{\infty}(-1)^k\frac{d^{(n)}}{dt^{(n)}}\frac{\partial L}{\partial (x^i)^{(n)}},
\end{equation}
where $(x^i)^{(n)}$ denotes the $n-$derivative with respect to time of the coordinate $x^i$.

The equations of motion of the Lagrangian (\ref{nonlocal-lag}) using the operator (\ref{EL-extend}) are precisely (\ref{eq-mot-inv}).
The second order dynamics (\ref{sec-or-dy}) is equivalent to the equations of motion (\ref{eq-mot-inv}) associated with the non-local Lagrangian (\ref{nonlocal-lag}), {\em i.e.},
\begin{eqnarray}
\label{sec-or-dy-1}
\nonumber
(1-C\frac{d}{dt})_{ij}\frac{\delta L_{NL}}{\delta x^j}=
A^{ij}\left(B_{jk}{\dot x}^k-\frac{\partial V}{\partial x^j}\right)\\
-\frac{d}{dt}\left[C_{ik}A^{kl}\frac{\partial V}{\partial x^l}\right]+\frac{d}{dt}\left[(  (A^T)_{il}+C_{ik}A^{kj}B_{jl}){\dot x}^l\right]=0
\end{eqnarray}
As the matrix operator $ (1-C\frac{d}{dt})_{ij}$ have a formal inverse (to obtain the momenta (\ref{p-j})) the system (\ref{sec-or-dy-1}) is equivalent with the system (\ref{sec-or-dy}).

\section{Noncommutative Classical Dynamics in configuration space}

In this section we will develop, as a simple application of the ideas above, the {\em noncommutative classical mechanics} constructed, as we stated in the introduction from a classical Hamiltonian function and a constant symplectic structure. We will outline some properties of this noncommutative dynamics  and the reduced non-local Lagrangian.

Lets start by the definition of the {\em noncommutative classical mechanics} as the dynamics described by the first order Lagrangian
\begin{equation}
\label{fo-lag-nc}
L=\alpha\left(\dot x p_x-y{\dot p}_y+\theta {\dot p}_y p_x-\sigma\dot x y\right)-H(x,y,p_x,p_y),
\end{equation}
where $\alpha=\frac{1}{1-\sigma\theta}$. In the case when $H=\frac12 p^2+V(x,y)$ the associated Lagrangian equations of motion are
\begin{eqnarray}
\label{eqpx}
\frac{\delta L}{\delta p_x}=\alpha(\dot x + \theta\dot{p_y})- p_x&=&0, \\
\label{eqpy}
\frac{\delta L}{\delta p_y}=\alpha(\dot y - \theta\dot{p_x})- p_y&=&0, \\
\label{eqx}
\frac{\delta L}{\delta x}=\alpha(-\dot{p_x} +\sigma\dot y)-\frac{\partial V}{\partial x}&=&0, \\
\label{eqy}
\frac{\delta L}{\delta y}=\alpha(-\dot{p_y} -\sigma\dot x)-\frac{\partial V}{\partial y}&=&0.
\end{eqnarray}
So the matrix $\sigma$ is
$$
\sigma=\alpha\begin{pmatrix}
0&\sigma&-1&0\\
 -\sigma&0&0&-1\\
1&0&0&\theta\\
0&1&-\theta&0
\end{pmatrix},
$$
from our previous notation   we can identify 
$$
A=\alpha\begin{pmatrix}
1&0\\ 0&1
\end{pmatrix}
,\quad B=\alpha\begin{pmatrix}
0&\sigma\\ -\sigma&0
\end{pmatrix}
, \quad C=\alpha\begin{pmatrix}
0&\theta\\ -\theta&0
\end{pmatrix}.
$$

The relation (\ref{GIS}) is in this case
\begin{eqnarray}\nonumber
 \frac{d}{dt}\left(\frac{\delta L}{\delta p_x}\right)
-\theta\frac{d}{dt}\left( \frac{\delta L}{\delta y} \right) 
 -\left( \frac{\delta L}{\delta x}\right) 
&=&\ddot{x}-\sigma\dot{y}+\frac{1}{\alpha}\frac{\partial V}{\partial x}-\theta\frac{d}{dt}(\frac{\partial V}{\partial y}) 
\cr \nonumber
\frac{d}{dt}\left(\frac{\delta L}{\delta p_y}\right)-\theta\frac{d}{dt}\left(\frac{\delta L}{\delta x} \right) -\left(\frac{\delta L}{\delta y}\right) &=&\ddot{y}+\sigma\dot{x}+\frac{1}{\alpha}\frac{\partial V}{\partial y}+\theta\frac{d}{dt}(\frac{\partial V}{\partial x}) .
\end{eqnarray}
From here we observe that there is a combination of equations of motion in ``phase space'' that can play the role of the ``definition of momenta''. The combination is
\begin{eqnarray}
 -\frac{\delta L}{\delta p_x}
-\theta \frac{\delta L}{\delta y}=0,\\
 -\frac{\delta L}{\delta p_y}
-\theta \frac{\delta L}{\delta x}=0.
\end{eqnarray}
Using these relations the system
\begin{eqnarray}
{\dot p}_x-\sigma\dot{y}+\frac{1}{\alpha}\frac{\partial V}{\partial x}=0, \\
{\dot p}_y+\sigma\dot{x}+\frac{1}{\alpha}\frac{\partial V}{\partial y}=0,\\
p_x=\dot x - \theta\frac{\partial V}{\partial x},\\
p_x=\dot y + \theta\frac{\partial V}{\partial y},
\end{eqnarray}
is equivalent to the original system.
The key point to notice here is that the definition of the momenta is not purely kinematics but contains dynamical information, {\em i.e.}, are different for each system defined by some potential function $V(x,y)$. This implies that the second order dynamics given by
\begin{eqnarray} 
\label{New-eq-x}
\ddot{x}-\sigma\dot{y}+\frac{1}{\alpha}\frac{\partial V}{\partial x}-\theta\frac{d}{dt}(\frac{\partial V}{\partial y})=0,\\
\label{New-eq-y}
\ddot{y}+\sigma\dot{x}+\frac{1}{\alpha}\frac{\partial V}{\partial y}+\theta\frac{d}{dt}(\frac{\partial V}{\partial x}) =0,
\end{eqnarray}
is not directly equivalent with the original first order system. We need to specify appropiarte boudary conditions to associate to a given solution of the original first order system the corresponding solution for the second order system (\ref{New-eq-x}, {New-eq-y}). For example the Lagrangian (\ref{fo-lag-nc}) implies the following  fixing at the boundaries
\begin{eqnarray}
\label{bound-1}
(x+\frac{\theta}{2}p_y)(t_i)=x_i,\quad (x+\frac{\theta}{2}p_y)(t_f)=x_f,\\
\label{bound-2}
(y-\frac{\theta}{2}p_x)(t_i)=y_i,\quad (y-\frac{\theta}{2}p_x)(t_f)=y_f.
\end{eqnarray}
For each trajectory with the specified boundary conditions (\ref{bound-1}) and (\ref{bound-2})
corresponds a solution of the system (\ref{New-eq-x}) and (\ref{New-eq-y}) with the boundaries
\begin{eqnarray}
\label{bound-New-1}
\Big(x+\frac{\theta}{2}(\dot y +\theta\frac{\partial V}{\partial x})\Big)(t_i)=x_i,\quad 
\Big(x+\frac{\theta}{2}(\dot y +\theta\frac{\partial V}{\partial x})\Big)(t_f)=x_f,\\
\label{bound-New-2}
\Big(y-\frac{\theta}{2}(\dot x -\theta\frac{\partial V}{\partial y})\Big)(t_i)=y_i,\quad 
\Big(y-\frac{\theta}{2}(\dot x -\theta\frac{\partial V}{\partial y})\Big)(t_f)=y_f.
\end{eqnarray}
This is the exact sense in which the original and the reduced second order dynamics are equivalent. It is worth noting that these boundary conditions depends on the parameter $\theta$ and on the specific potential that appears in the Newtonian equations (\ref{New-eq-x}) and (\ref{New-eq-y}). This information is crucial for the quatization of the classical system using the Path Integral.

To construct the  non-Local Lagrangian associated to the equations of motion  (\ref{New-eq-x}) and (\ref{New-eq-y}) we obtain the momenta from equations (\ref{eqpx}) and (\ref{eqpy}) to get
\begin{equation}
p_x=\alpha(1+(\alpha\theta)^2\frac{d^2}{dt^2})^{-1}\frac{d}{dt}(x+\alpha\theta\dot y), \quad
p_y=\alpha(1+(\alpha\theta)^2\frac{d^2}{dt^2})^{-1}\frac{d}{dt}(y-\alpha\theta\dot x).
\end{equation}

According to our analysis the equations of motion in the noncommutative configuration space are
\begin{eqnarray}
\label{non-local-1}
\frac{1}{\alpha^2}\frac{\partial V}{\partial x}-\frac{\sigma}{\alpha}\dot y+ (1+(\alpha\theta)^2 \frac{d^2}{dt^2})^{-1}\frac{d^2}{dt^2}(x+\alpha\theta \dot y)=0,\\
\label{non-local-2}
\frac{1}{\alpha^2}\frac{\partial V}{\partial y}+\frac{\sigma}{\alpha}\dot x+ (1+(\alpha\theta)^2 \frac{d^2}{dt^2})^{-1}\frac{d^2}{dt^2}(y-\alpha\theta \dot x)=0,
\end{eqnarray}   
and the Lagrangian is
\begin{equation}
L=\frac12(\dot x\star\dot x+\dot y\star\dot y)+\frac{\alpha\theta}{2}(\dot x\star\ddot y- 
\dot y\star\ddot x)+\frac{\alpha\sigma}{2}(x\dot y-y\dot x)-V(x,y),
\end{equation}
where
\begin{equation}
\star\equiv\alpha^2(1+(\alpha\theta)^2\frac{d^2}{dt^2})^{-1}.
\end{equation}

The equations (\ref{non-local-1}) and (\ref{non-local-2}) are equivalent to the second order equations  (\ref{New-eq-x}) and (\ref{New-eq-y}).

\section{Examples}

Consider the following Lagrangian \cite{Hojman-Shepley},
$$
L=(y+z)\dot x + w\dot z + \frac12 (w^2-2zy-z^2).
$$
The symplectic structure that can be read out from this Lagrangian is
$$
\{x,y\}=1,\qquad \{y,w\}=-1,\qquad \{w,z\}=-1,
$$
(all other Poisson brackets are zero). The equations of motion are
\begin{eqnarray}
\label{x}
\frac{\delta L}{\delta x}= -\dot y-\dot z=0,\\
\label{y}
\frac{\delta L}{\delta y}=\dot x - z=0,\\
\label{w}
\frac{\delta L}{\delta w}=\dot z+w=0,\\
\label{z}
\frac{\delta L}{\delta z}=\dot x - \dot w-y-z=0.
\end{eqnarray}
At first sight the variables $y,z$ as a set and $w$ are auxiliary. In fact, using (\ref{y}),(\ref{z}) and (\ref{w}) we can define a reduced Lagrangian for the variable $x$ that gives the correct reduced Lagrangian and four order equation of motion for $x$. We can also consider as auxiliary variables only $y,z$ leaving the reduced space for $x,w$ whose Lagrangian can also be constructed. In this case we observe that the reduced coordinates define a commutative configuration space ($x,w$ have zero Poisson bracket).  

Now suppose that we want to construct the reduced dynamics in the noncommutative space $x,y$. To arrive to this space we can use the equations (\ref{y}) and (\ref{w}) to eliminate the variables $z,w$ but this reduction is not variational admissible  so we can not construct a Lagrangian for this reduced system. In fact, it can be shown that a quadratic Lagrangian for this reduced second order dynamics does not exist.

The construction of the non-local Lagrangian can be worked out by 
the procedure sketched in this note and  consist in the use of the equations of motion (\ref{z}) and (\ref{w}) to obtain the variables $z,w$ from its own equations of motion. The first observation is that these variables are not auxiliary in the usual sense, but nevertheless we can obtain them as a formal integration of the equations (\ref{z}) and (\ref{w}) to get
\begin{equation}
z=(1-\frac{d^2}{dt^2})^{-1}(\dot x -y),\qquad
w=-(1-\frac{d^2}{dt^2})^{-1} \frac{d}{dt}(\dot x -y).
\end{equation}
The reduced equations of motion are
\begin{equation}
\label{lag-eq-mot}
\dot x -(1-\frac{d^2}{dt^2})^{-1}(\dot x -y)=0, \qquad
\dot y +(1-\frac{d^2}{dt^2})^{-1}\frac{d}{dt}(\dot x -y)=0.
\end{equation}
These equations are equivalent to the second order equations
\begin{equation}
\label{red-sys}
\ddot y + y=0,\qquad \ddot x+\dot y=0.
\end{equation}
The reduction process, namely the  construction of a Lagrangian associated to this equations of motion can be implemented to get
\begin{equation}
L_R= \frac12 (\dot x - y)(1-\frac{d^2}{dt^2})^{-1}(\dot x - y)+y\dot x. 
\end{equation}
This Lagrangian reproduces the equations of motion (\ref{lag-eq-mot}). 
The relation between the Lagrangian equations of motion and the reduced second order dynamics is given by
\begin{eqnarray}
\frac{d}{dt}\frac{\delta L}{\delta y}+\frac{\delta L}{\delta x}=\ddot x + \dot y,\\
 -\frac{d}{dt}\frac{\delta L}{\delta x}-\frac{d}{dt}\frac{\delta L}{\delta w}-
\frac{\delta L}{\delta z}+\frac{\delta L}{\delta y} =\ddot y + y.
\end{eqnarray}
From here we can see that the combination $\frac{\delta L}{\delta x}+\frac{\delta L}{\delta w}=0$ and $\frac{\delta L}{\delta y}=0$ define the ``momenta''
\begin{equation}
\label{z-w-momenta}
w-\dot y=0,\quad \dot x-z=0.
\end{equation}
The system
\begin{equation}
w-\dot y=0,\quad \dot x-z=0,\quad \dot z+w=0,\quad \dot x - \dot w-y-z=0,
\end{equation}
is equivalent to the original set of first order equations of motion. In this case the general solution of the original first order system and the general solution for the reduced system (\ref{red-sys}) coincides because the ``momenta'' $z,w$ as given by (\ref{z-w-momenta}) are precisely the velocities as in the standard Newtonian case.

The case of the noncommutative classical mechanics analyzed in section 5 can be illustrated by taking as an example the simple potential $V=c(x+y)$ with $c$ any constant. Here the general solutions of the original first order system and the reduced one do not coincide, but taking into account the properly choosed boundary conditions (\ref{bound-New-1}) and (\ref{bound-New-2}) the respective solution coincides and are
\begin{eqnarray}
\nonumber
x(t)=\frac{1}{4T}\Big(-4t_f x_i+4t_i x_f +4 t X
 -2 \theta Y-\\
\nonumber
cT\Big(2t^2+\theta(t_f+\theta)-2t(t_i+t_f+\theta)
+ t_i(2t_f+\theta)\Big)\Big)\\
\nonumber
y(t)=\frac{1}{4T}\Big(-4t_f y_i+4t_i y_f +4 t Y
 +2 \theta X-\\
\nonumber
cT\Big(2t^2+\theta(\theta-t_f)-2t(t_i+t_f-\theta)
+ t_i(2t_f-\theta)\Big)\Big).
\end{eqnarray}
where $X=x_i-x_f, Y=y_i-y_f,T=t_i-t_f$. The free particle can be obtained by taking $c=0$ in the expressions above. It is interesting to observe that due to the boundary conditions the solution depends on the parameter $\theta$.

As a final example consider the construction of a second order Lagrangian for the noncommutative harmonic osillator described by the first order Lagrangian (\ref{fo-lag-nc}), with the potential given by $V=\frac12\omega^2(x^2+y^2)$, using a Darboux transformation to render some  variables to auxiliary variables. The Darboux transformation is
\begin{equation}
\label{Darboux}
x'=x+\frac12\theta p_y, \quad y'=y-\frac12 \theta p_x, \quad  p_x'=p_x,\quad p_y'=p_y.
\end{equation}
With this transformation we construct a new first order Lagrangian $L'$. For this Lagrangian the momenta $p_x',p_y'$ are auxiliary variables. Then we proceed to eliminate them to find the second order Lagrangian
$$L'=\Omega^2\left[({\dot x}'^2+{\dot y}'^2)-\frac12\omega^2(x'^2+y'^2)+\frac{\theta\omega^2}{2}(x'{\dot y}'-y'{\dot x}')\right]$$
where $\Omega=(1+(\frac{\omega\theta}{2})^2)^{-1}$.
The crucial point is that the equations of motion associated to this second order Lagrangian are  equivalet to the original second order system. The equations of motion associated with this Lagrangian are
$$\Omega^2\left[{\ddot x}' +\omega^2 x' +\omega^2\theta{\dot y}'\right]=0,$$
$$\Omega^2\left[{\ddot y}' +\omega^2 y' +\omega^2\theta{\dot x}'\right]=0.$$
These equations of motion are proportional to the original system of second order equations. So the Darboux transformation (\ref{Darboux}) is a symmetry of the equations of motion. So, we have constructed an equivalent Lagrangian for the given system of second order equations. The symmetry transformation is given by the projection of the Darboux transformation to the configuration space
$$
x=x'-\frac12\theta\Omega^2({\dot y}'+\frac{\omega^2\theta}{2}x'), \quad
y=y'-\frac12\theta\Omega^2({\dot x}'+\frac{\omega^2\theta}{2}y').
$$

\section{Conclusions}

In summary, we have analyzed the role of auxiliary variables in first order dynamics and its relation with the inverse problem of the calculus of variations. The relation between the first order equations and the reduced second order equations was worked out by showing that there exist a relation between a functional combination of the first order equations of motion and the reduced system. This relation can always be computed when the variables of the reduced space are fixed and works a the level of the equations of motion. When the eliminated variables are auxiliary, the Lagrangian can be constructed directly. When the variables are not auxiliary we need to find a symmetry of the equations of motion that allows the construction of a Lagrangian with auxiliary variables. In this way we can construct an equivalent second order Lagrangian for a given set of second order equations of motion.

When the symplectic structure is constant we can construct, in a formal way, a non-local Lagrangian. The equations of motion of this non-local Lagrangian are equivalent to the set of second order equations (\ref{sec-or-dy}). These equations are in turn equivalent  to the original  set of first order equations of motion.

Finally we have worked out as examples the case of ``non-commutaive mechanics'', a system with no second order Lagrangian and for the case of the non-commutaive harmonic oscillator we have constructed using our procedure an equivalent Lagrangian for the original second order dynamics. It will be interesting to analyze the problem of the quantization \cite{Hojman-Shepley} of this systems and its relation with the non-commutaive quantum mechanics.

\section*{Acknowlegments} 

This work was supported in part by grants CONACyT 32431-E and DGAPA IN11700.
We would like to thank to Marc Henneaux for calling our attention to the reference \cite{Henneaux} where the same perspective to study the Inverse problem of the calculus of variation was adopted.

\end{document}